\documentclass[prl,showpacs,byrevtex,floatfix,twocolumn,superscriptaddress]{revtex4}
\usepackage{amssymb,graphicx,afterpage,amsmath,color}
\begin{document}
\title{\boldmath Giant directional dichroism of terahertz light in resonance with magnetic
excitations of the multiferroic oxide BaCo$_2$Ge$_2$O$_7$ \unboldmath}

%
%
%
\author{I. K\'ezsm\'arki}
\affiliation{Multiferroics Project, ERATO, Japan Science and Technology Agency (JST), Japan c/o The University of Tokyo, Tokyo 113-8656, Japan} \affiliation{Department of Physics, Budapest University of
Technology and Economics and Condensed Matter Research Group of the Hungarian Academy of Sciences, 1111 Budapest, Hungary}
\author{N. Kida}
\affiliation{Multiferroics Project, ERATO, Japan Science and Technology Agency (JST), Japan c/o The University of Tokyo, Tokyo 113-8656, Japan}
\author{H. Murakawa}
\affiliation{Multiferroics Project, ERATO, Japan Science and Technology Agency (JST), Japan c/o The University of Tokyo, Tokyo 113-8656, Japan}
\author{S. Bord\'acs}
\affiliation{Multiferroics Project, ERATO, Japan Science and Technology Agency (JST), Japan c/o The University of Tokyo, Tokyo 113-8656, Japan} \affiliation{Department of Physics, Budapest University of
Technology and Economics and Condensed Matter Research Group of the Hungarian Academy of Sciences, 1111 Budapest, Hungary}
\author{Y. Onose}
\affiliation{Multiferroics Project, ERATO, Japan Science and Technology Agency (JST), Japan c/o The University of Tokyo, Tokyo 113-8656, Japan}
\affiliation{Department of Applied Physics, University of Tokyo,
Tokyo 113-8656, Japan}
\author{Y. Tokura}
\affiliation{Multiferroics Project, ERATO, Japan Science and Technology Agency (JST), Japan c/o The University of Tokyo, Tokyo 113-8656, Japan}
\affiliation{Department of Applied Physics, University of Tokyo,
Tokyo 113-8656, Japan}
\affiliation{Cross-correlated materials group (CMRG) and correlation electron research group
(CERG), RIKEN Advanced Science Institute, Wako 351-0198, Japan}
\date{\today}
\begin{abstract}
We propose that concurrently magnetic and ferroelectric, i.e. multiferroic, compounds endowed with electrically-active magnetic excitations (electromagnons) provide a key to produce large directional dichroism for long wavelengths of light. By exploiting the control of ferroelectric polarization and magnetization in a multiferroic oxide Ba$_2$CoGe$_2$O$_7$, we demonstrate the realization of such a directional light-switch function at terahertz frequecies in resonance with the electromagnon absorption. Our results imply that this hidden potential is present in a broad variety of multiferroics.
\end{abstract}
\maketitle

Matter can change color when exposed to a magnetic field. The most exotic manifestation of such magneto-chroism is the non-reciprocal directional dichroism, that is, materials can distinguish between counter-propagating light beams irrespective of their light polarization. Such a violation of reciprocity can be manifested not only in absorption but also in scattering and refraction processes each governed by the difference in the light velocity (complex index of refraction) for forward and backward propagation \cite{O'Dell1970,Barron2004}. The symmetry requirements of directional anisotropy --sharing common origin with the dc magnetoelectric effect \cite{Brown1963}-- are rather strict as time reversal and spatial inversion should be simultaneously broken \cite{Baranova1977,Fiebig2005}. Concerning the microscopic nature of the light-matter interaction, it is closely related to the magneto-electric linear dichroism, the Jones effect, and the magneto-chiral dichroism \cite{Rikken2000,Rikken2002,Barron2004}. Despite the early prediction of this phenomenon by Brown and coworkers \cite{Brown1963}, and the efforts donated to its detection with reducing the symmetry of matter, e.g. by application of crossed static electric and magnetic field, the effect was usually found weak \cite{Krichevtsov1993,Rikken2002_2,Jung2004}. Promising hosts of this phenomenon could be multiferroic compounds and metamaterials where both the spontaneous magnetization and ferroelectric polarization are built in the media and may produce large amplification \cite{Arima2008,Kida2006}. The existence of directional anisotropy was confirmed so far in the near infrared, visible and ultraviolet region of the spectrum and, more recently, for X-ray radiation \cite{Kubota2004}. However, its fingerprints have never been evidenced for long wavelengths of light covering the range of magnetic excitations and lattice vibrations in solids. Based on the general concept of multiferroic systems, we show that spin excitations when coupled to electric polarization can be particularly effective to generate directional dichroism controllable by moderate magnetic fields.

In multiferroics, the magnetic and polarization order are coupled to produce the dc magnetoelectric effect; electric polarization can be induced by magnetic field and magnetization appears in response to an electric field \cite{Kimura2003,Fiebig2005,Tokura2006,Wang2009}. If the ground state has such an entanglement, the low-lying excitations must also be interlocked. Thus, oscillating magnetization and polarization ($M_i^\omega$ and $P_i^\omega$) can be induced by the electric and magnetic component of light ($E_i^\omega$ and $H_i^\omega$), respectively, through the optical magnetoelectric effect (OME) according to $M_i^\omega$$=$$4\pi\chi^{me}_{ij}(\omega)E_j^\omega$ and $P_i^\omega$$=$$4\pi\chi^{em}_{ij}(\omega)H_j^\omega$. Spin and polarization waves can hybridize to form \emph{electromagnons}, as discussed e.g. in RMnO$_3$ \cite{Pimenov2006}, RMn$_2$O$_5$ \cite{Sushkov2007}, and Ba$_2$Mg$_2$Fe$_{12}$O$_{22}$ \cite{Kida2009}. A microscopic view based on the Kubo formula shows that resonant electromagnon processes can most efficiently generate the OME:
\begin{equation}
\chi^{me}_{ij}(\omega)=
\frac{2}{\hbar}\sum_n
\frac{\omega_{no}\Re\{\langle 0|m_i|n\rangle\langle n|e_j|0\rangle\}+i\omega\Im\{ \langle 0|m_i|n\rangle \langle n|e_j|0\rangle\}}{\omega_{no}^2-\omega^2-2i\omega\delta}.
\label{Kubo}
\end{equation}
Dominant contributions arise when transitions from the ground state
$|0\rangle$ to excited states $|n\rangle$ can be resonantly induced by photons, i.e. $\omega\thickapprox\omega_{no}$, both via magnetic ($m_i$) and electric ($e_j$) dipole operators. Terms associated with the real ($^\prime$) / imaginary ($^{\prime\prime}$) part of the transition matrix elements change sign / remain invariant under time reversal, respectively. The two cross-effects are related by the Kubo formula according to $\chi^{me}_{ij}(\omega)$$=$$\chi^{\prime}_{ij}(\omega)$$+$$\chi^{\prime\prime}_{ij}(\omega)$ and $\chi^{em}_{ij}(\omega)$$=$$\chi^{\prime}_{ji}(\omega)$$-$$\chi^{\prime\prime}_{ji}(\omega)$. As also obvious from Eq.~(\ref{Kubo}) the OME changes sign under spatial inversion.

For the study of directional dichroism via the OME of electromagnon excitations, the multiferroic phase of Ba$_2$CoGe$_2$O$_7$ offers an ideal arena. A representative part of its tetragonal crystal structure (space group P$\overline{4}$2$_1$m) is shown in Fig.~1a. At room temperature it is a paramagnetic insulator with a charge gap of $\Delta$$\approx$$4$\,eV. According to preceding study of neutron scattering \cite{Zheludev2003} and magnetization \cite{Yi2008,Murakawa,Sato2003}, below T$_N$$=$$6.7$\,K the spin of Co$^{2+}$ ions lie in the tetragonal plane and form an antiferromagnetic structure with tiny canting along the [110] (or [1$\overline{1}$0]) direction.
The onset of the spontaneous magnetization, \textbf{M}, is accompanied with a ferroelectric polarization, \textbf{P}, pointing along the tetragonal [001] axis \cite{Yi2008,Murakawa} resulting in a ferrotoroidic state; \textbf{T}$=$\textbf{P}$\times$\textbf{M} \cite{Fiebig2005}. The field dependence of the magnetization and the polarization together with the spin texture can be discerned in Fig.~1b-c. (Hereafter, the crystallographic axes [110], [1$\overline{1}$0] and [001] will be referred to as \textbf{x}, \textbf{y} and \textbf{z} direction, respectively.)
Though the polarization is an even function of the magnetic field, it can be alternatively reversed either by $\pi/2$ rotation of the field within the $xy$-plane \cite{Murakawa} or by $\pi$ rotation of the sample around the \textbf{y} axis, as shown in Fig.~1d. The feature that \textbf{M} and \textbf{P} can be separately switched turns out to be a key factor in the control of the directional dichroism.
\begin{figure}[t!]
\includegraphics[width=3.2in]{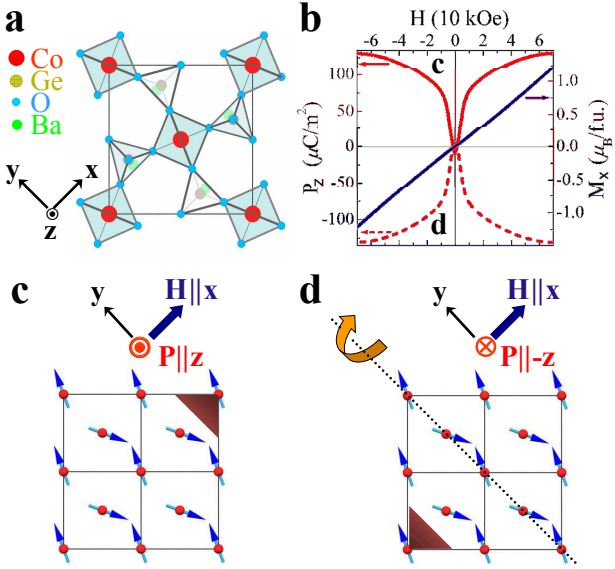}
\caption{\textbf{a,} Lattice structure of Ba$_2$CoGe$_2$O$_7$ within the layer perpendicular to the tetragonal axis, \textbf{z}. Magnetic Co$^{2+}$ ions with S=3/2 spin form square lattice. \textbf{b,} The magnetic field (\textbf{H}$\parallel$\textbf{x}) dependence of the magnetization (M$_x$) and the polarization (P$_z$) at $T$$=$$3.4$\,K. Note that the polarization is an even function of $H$. \textbf{c,} In magnetic fields parallel to the \textbf{x} axis a single domain is formed with the polarization pointing to the \textbf{z} direction. The approximate spin structure as determined by Ref.~\onlinecite{Zheludev2003} is also indicated. \textbf{d,} By $\pi$ rotation of the crystal around the \textbf{y} axis, the polarization is reversed in the frame of the observer while the magnetization is fixed by the external field (see also the panel \textbf{b}).}
\label{fig1}
\end{figure}

In the multiferroic phase sketched in Fig.~1c-d, the symmetry is lowered to orthorhombic with a magnetic point group of m$_x$m$^\prime_y$2$^\prime_z$ which corresponds to that of an originally isotropic media after introducing crossed static electric and magnetic field \cite{Rikken2002_2,Fiebig2005,Wang2009}. However, as we will demonstrate, the spontaneous symmetry-breaking fields are far more powerful in producing the OME. Neumann's principle yields the following form for the magnetoelectric tensor in accordance with field-dependent electric polarization measurements \cite{Yi2008,Murakawa}:
\begin{equation}
\chi_{ij}^{me}=\left[
\begin{array}{ccc}
0 & \chi^{\prime\prime}_{xy} & \chi^{\prime}_{xz} \\
\chi^{\prime\prime}_{yx} & 0 & 0 \\
\chi^{\prime}_{zx} & 0 & 0
\end{array}
\right].
\label{tensor}
\end{equation}
\begin{figure*}[t!]
\includegraphics[width=4.9in]{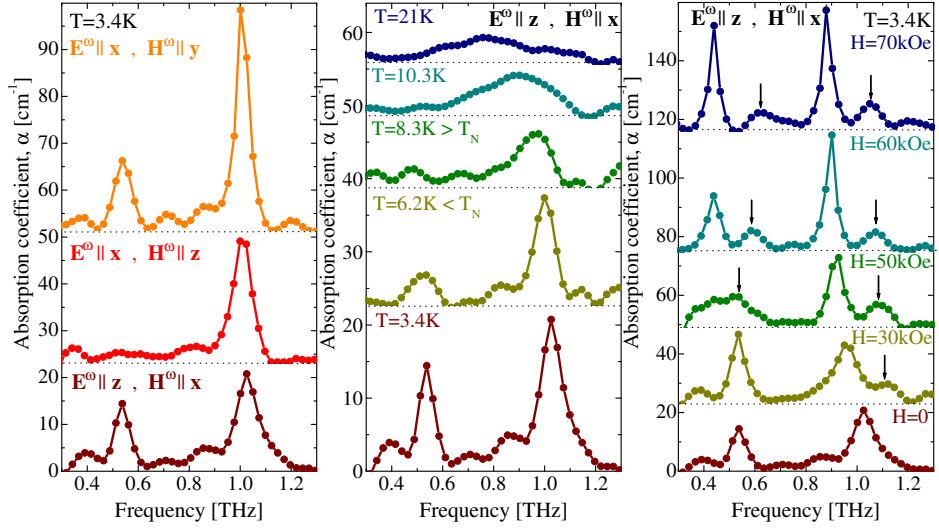}
\caption{Selection rules and characteristics for the magnon modes. The spectra are shifted relative to each other with the corresponding baselines. \textbf{a,} Light polarization dependence at $T$$=$$3.4$\,K in zero field when the \textbf{x} and \textbf{y} directions are equivalent. The $0.5$\,THz mode is excited by the $\textbf{H}^\omega$ magnetic component of light but insensitive to the change of the $\textbf{E}^\omega$ electric component. By contrast, the $1$\,THz mode is clearly affected by the variation of either $\textbf{H}^\omega$ or $\textbf{E}^\omega$. \textbf{b,} Temperature dependence of the absorption spectrum. The conventional magnon disappears above the N\'eel temperature, while the electromagnon at $1$\,THz subsists well above T$_N$ with a considerable fraction of its oscillatory strength present even at $\thicksim20$\,K. \textbf{c,} At $T$$=$$3.4$\,K the both modes split into two excitations (the side peaks are indicated by arrows) with the $g$ factor $\sim$2 in high fields.}
\label{fig2}
\end{figure*}

Using terahertz time-domain spectroscopy we have investigated the spin excitations in high-quality single crystals of Ba$_2$CoGe$_2$O$_7$ with typical thickness of $1$\,mm over the frequency range of $0.2-2$\,THz (0.8-8meV). (Detailed description of the sample preparation and the applied terahertz method are given in Ref.~\onlinecite{Murakawa} and \onlinecite{Kida2008}, respectively.) Besides the determination of the temperature and field dependence of the transition energies, we have performed a systematic study with change of light polarization, $\textbf{E}^\omega$ and $\textbf{H}^\omega$, to establish the selection rules for the different magnon branches. In the multiferroic phase we observed two distinct absorption bands located at $\thicksim0.5$\,THz and $\thicksim1$\,THz in zero field (see Fig.~2a). While the higher-energy mode can be excited for any polarization configuration, the lower one is silent if the magnetic component of the light is parallel to the tetragonal axis, \textbf{H}$^\omega$$\parallel$\textbf{z}. The strength of the 0.5\,THz band is independent of the orientation of the light $\textbf{E}^\omega$ vector implying that this excitation has a dominantly magnetic-dipole character. On the other hand, as it is clear from the main polarization configurations shown in Fig.~2a, the intensity of the sharp transition at $\thicksim1$\,THz changes if either $\textbf{H}^\omega$ or $\textbf{E}^\omega$ is varied. Thus, this mode is an \emph{electromagnon} with both electric- and magnetic-dipole activity, and hence should resonantly mediate the OME as expected from Eqs.~(\ref{Kubo})-(\ref{tensor}). From now on, we focus on the configuration of \textbf{E}$^\omega$$\parallel$\textbf{P} ($\parallel$\textbf{z}) $\&$ \textbf{H}$^\omega$$\parallel$\textbf{M} ($\parallel$\textbf{x}), since it turns out to be a unique condition for the directional dichroism phenomenon.

As expected for spin waves, with increasing temperature the absorption of the $0.5$\,THz mode is gradually reduced and disappears above $T_N$$\thickapprox$$6.7$\,K (see Fig.~2a). In contrast, the electromagnon at $1$\,THz survives above $T_N$ with an increasing red-shift and broadening. Both modes show conspicuous magneto-chroism; their intensity is enhanced with the application of magnetic field along the \textbf{x} direction and a splitting with a g-factor $\thicksim2$ in high fields can be also followed in Fig.~2c.

We turn to the extraordinary aspect of the electromagnon, namely its strong directional dichroism. The solution of the Maxwell equations for the symmetry of the multiferroic Ba$_2$CoGe$_2$O$_7$ gives clear guidelines. Only when the light propagates perpendicular to both \textbf{M} and \textbf{P}, i.e. \textbf{k}$=$$(0,k_y,0)$, the directional degeneracy of the optical response is lifted and the velocity becomes different for $\pm k_y$. Consequently, we obtain four independent solutions for the generalized complex refractive index each corresponding to linearly polarized eigenstates:
\begin{eqnarray*}
N^{\parallel}_{\pm}(\omega)\approx\pm4\pi\chi_{xz}^{\prime}(\omega)+\sqrt{\varepsilon_{zz}(\omega)\mu_{xx}(\omega)}\\
N^{\perp}_{\pm}(\omega)\approx\pm4\pi\chi_{zx}^{\prime}(\omega)+\sqrt{\varepsilon_{xx}(\omega)\mu_{zz}(\omega)},
\end{eqnarray*}
where $\varepsilon_{ij}$ and $\mu_{ij}$ are elements of the electric permittivity and magnetic permeability tensors. $N^{\parallel}_{\pm}$ is the refractive index for $\pm k_y$ in case \textbf{E}$^\omega$ is parallel to the ferroelectric polarization and \textbf{H}$^\omega$ points along the magnetization, while $N^{\perp}_{\pm}$ stands for the perpendicular polarization (\textbf{E}$^\omega$$\parallel$\textbf{M} $\&$ \textbf{H}$^\omega$$\parallel$\textbf{P}). Then the non-reciprocal term in the complex refractive index, whose real/imaginary part corresponds to the directional birefringence/dichrosim, is obtained as $\Delta N^{\parallel}$$=$$N^{\parallel}_{+}-N^{\parallel}_{-}$$=$$8\pi\chi_{xz}^{\prime}$ and $\Delta N^{\perp}$$=$$8\pi\chi_{zx}^{\prime}$. Since $\Delta N$ changes sign either by time reversal or spatial inversion, the switching of the magnetization or the polarization of the material is equivalent to the reversal of the light propagation. Therefore, the capability to independently control \textbf{M} and \textbf{P} is crucial for the rigorous analysis of the optical magnetoelectric phenomena.
\begin{figure}[h!]
\includegraphics[width=3.5in]{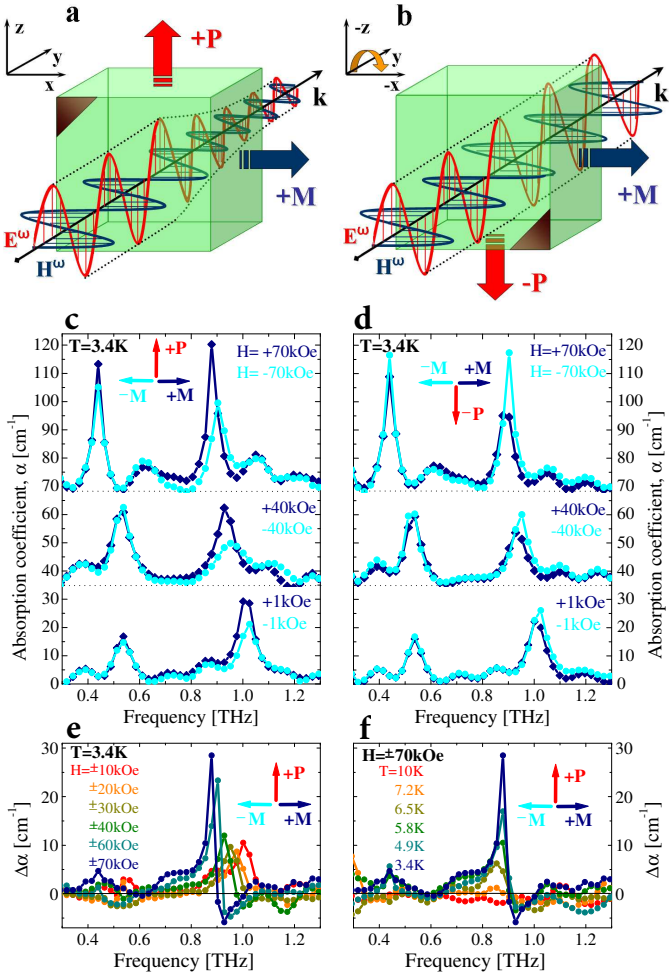}
\caption{Directional dichroism, i.e. the change in the absorption
upon reversal of the light propagation, can be equivalently induced by the switching of either the static magnetization (\textbf{M}) or the polarization (\textbf{P}) of the sample. \textbf{a,} Due to the resonant absorption the radiation decays as passing through the material. This loss can be controlled both by \textbf{M}\ ($\parallel\ $$\textbf{H}^\omega$) and \textbf{P}\ ($\parallel\ $$\textbf{E}^\omega$). \textbf{b,} For example, the reversal of \textbf{P} by $\pi$ rotation of the sample around the \textbf{y} axis can reduce the absorption which is somewhat exaggerated for clarity. \textbf{c \& d,} Only the absorption of
the electromagnon has a large component odd both in \textbf{M} and \textbf{P}, thus shows
directional dichroism, $\Delta\alpha$. \textbf{e,} As the magnetic field is
increased up to H=70\,kOe, $\Delta\alpha$ spectra grows monotonously and the peak position exhibits a red-shift. \textbf{f,} The OME vanishes with the loss of long-range order above T$_N$=6.7\,K even in H=70\,kOe.}
\label{fig3}
\end{figure}

Fig.~3a-b schematically represent the configuration in which we probed the existence of directional dichroism. We found large non-reciprocal absorption reaching $\Delta\alpha$$\approx$$30$\,cm$^{-1}$ in the vicinity of the electromagnon at 1\,THz -- corresponding in maximum to a relative change of the absorption coefficient $\Delta\alpha/\alpha_{ave}$$=$$2(\alpha^{+M}-\alpha^{-M})/(\alpha^{+M}+\alpha^{-M})$$\approx$$0.6$ in $H$$=$$\pm70$\,kOe -- but no such effect for the magnon mode at $0.5$\,THz as discerned in Fig.~3c-d. Furthermore, $\Delta\alpha$ is reversed when the ferroelectric polarization is "switched" by $\pi$ rotation of the sample around the \textbf{y} axis as shown in Fig.~3b. Fig.~3c-d demonstrate that in easily accessible fields ($H$$\sim$$\pm1$\,kOe) the effect remains as high as $\Delta\alpha/\alpha_{ave}$$\approx$$0.2$. Its further increase with the field strength is in accord with the evolution of the symmetry-breaking order parameters \textbf{M} and \textbf{P} (see Fig.~3e). While the absorption by the electromagnon mode subsists to some extent above $T_N$, the temperature dependence of $\Delta\alpha$ in Fig.~3f shows that the giant OME is restricted to the multiferroic state.

As argued above, the observation of directional dichroism allows the direct measurement of the magnetoelectric tensor elements, especially those odd in magnetic field. Using the relation between the refractive index and the absorption coefficient, $\Delta\alpha^{\parallel}$$=$$2\omega/c\cdot\Im\{\Delta N^{\parallel}\}$$=$$16\pi\omega/c\cdot\Im\{\chi^{\prime}_{xz}\}$, we get $4\pi\Im\{\chi_{xz}^{\prime}\}$$\approx$$4\times10^{-2}$ for the electromagnon at $T$$=$$3.4$\,K in $H$$=$$70$\,kOe. This number may look small when compared with typical values of the dielectric constant for electric-dipole transitions but it is appreciable on the scale of magnetic-dipole processes. Indeed, for the $0.5$\,THz magnon excitation --using the background dielectric constant $\varepsilon_{zz}$$\approx$$8$ determined from THz transmission data-- we obtain $\Im\{\mu_{xx}\}$$\approx$$1.2\times10^{-1}$ under the same conditions. Thus, the OME generating the electromagnon mode at $1$\,THz is the same order of magnitude as the magnetic susceptibility associated with the pure magnon band. In fact, $\Delta\alpha/\alpha_{ave}$ found in the order of unity means that the imaginary part of the magnetoelectric susceptibility around $\omega$$\approx$$1$\,THz is close to the upper theoretical bound allowed by the second law of thermodynamics according to $\Im\{\sqrt{\varepsilon_{zz}(\omega)\mu_{xx}(\omega)}-4\pi\chi_{xz}^{\prime}(\omega)\}$$\geq$$0$. Though the symmetry would also allow directional dichroism for the other polarization when \textbf{E}$^\omega$$\parallel$\textbf{M} $\&$ \textbf{H}$^\omega$$\parallel$\textbf{P}, we found that $\Delta\alpha^{\perp}$$\approx$$0$ within the precision of the experiment.

In conclusion, we showed that in multiferroic materials the OME can be resonantly enhanced in the range of spin excitations through an electric dipole-magnetic dipole interference. Such electromagnon excitations give a unique opportunity to fabricate magnetically controllable directional light-switch for unpolarized microwave and terahertz radiation based on bulk materials. We demonstrated for the case of multiferroic Ba$_2$CoGe$_2$O$_7$ that this idea indeed works and a directional dichroism in the order of unity can be realized within the THz-region of the electromagnetic spectrum.

We thank to R. Shimano, T. Arima, S. Miyahara, N. Nagaosa, A. J\'anossy, K. Penc, and G. Mih\'aly  for discussions. This work was supported by KAKENHI, MEXT of Japan, by Funding Program for World-Leading Innovation R\&D on Science and Technology (FIRST) on "Strong-correlation quantum science", and by Hungarian Research Funds OTKA PD75615, NK72916, Bolyai 00256/08/11, T\'AMOP-4.2.1/B-09/1/KMR-2010-0002.

\end{document}